# Dynamic Global Constraints: A First View[+]

Roman Barták[*]

Charles University
Faculty of Mathematics and Physics
Malostranské náměstí 2/25, Praha 1, Czech Republic
bartak@kti.mff.cuni.cz

**Abstract.** Global constraints proved themselves to be an efficient tool for modelling and solving large-scale real-life combinatorial problems. They encapsulate a set of binary constraints and using global reasoning about this set they filter the domains of involved variables better than arc consistency among the set of binary constraints. Moreover, global constraints exploit semantic information to achieve more efficient filtering than generalised consistency algorithms for n-ary constraints. Continued expansion of constraint programming (CP) to various application areas brings new challenges for design of global constraints. In particular, application of CP to advanced planning and scheduling (APS) requires dynamic additions of new variables and constraints during the process of constraint satisfaction and, thus, it would be helpful if the global constraints could adopt new variables. In the paper, we give a motivation for such dynamic global constraints and we describe a dynamic version of the well-known `alldifferent` constraint.

## 1 Introduction

Constraint programming (CP) is successfully applied to real-life combinatorial (optimisation) problems thanks to its declarative character, which supports natural modelling of real-life problems, and thanks to general solving technology, which can encapsulate various solving techniques. One of the main features of CP is locality of the problem description i.e. we model the problem using a set of constraints and most of the constraints bind a small set of variables which is much smaller than the set of all variables in the problem. This is different from the traditional operation research (OR) approach where the "constraints" bind all the variables and thus, OR solving methods can exploit global reasoning about the problem. On the other hand, global

---

[+] This is a revised version of the paper presented at CP-AI-OR2001 Workshop.
[*] Supported by the Grant Agency of the Czech Republic under the contract no. 201/01/0942 and partially supported by the project LN00A056 of the Ministry of Education of the Czech Republic.

reasoning can be very expensive and a simple method of constraint propagation, pruning, filtering or whatever you call it, can reduce the search space more efficiently.

The goal of constraint propagation is to achieve some level of consistency in the network of constraints and variables by removing inconsistent values from variables' domains (inconsistent values cannot take part in any solution). Achieving a higher level of consistency leads to removing more inconsistent values but it is more expensive in terms of time and space. Consequently, a simple arc consistency or its generalised version for n-ary constraints is usually used. Application of CP to many real-life problems shows that a good trade-off between efficiency and level of consistency can be achieved by using global constraints instead of the sets of binary constraints [7].

A global constraint encapsulates several simple constraints and by exploiting semantic information about this set of constraints it can achieve better pruning of domains. Filtering algorithms for global constraints are based on methods of graph theory, discrete mathematics, or operation research so they make the bridge between these mathematical areas and search-based constraint programming with origins in artificial intelligence. However, the traditional global constraints require all the constrained variables to be known before the constraint is posted to the system. This feature makes using of global constraints more complicated in areas where new variables are generated during the course of solving. This difficulty can be solved using dummy variables that represent slots for variables to come. Unfortunately, this approach cannot be used if large number of dummy variables is required because it decreases efficiency of the filtering algorithm and it increases memory consumption.

In Section 2, we give some real-life motivated examples from advanced planning and scheduling to justify adding new variables during the course of problem solving. To remove difficulties with the dummy variables we propose a concept of dynamic global constraints that can adopt new variables during solving. In Section 3, we introduce a monotonic feature of the global constraints and then we describe how arbitrary monotonic global constraint can be adapted to a dynamic environment. This approach is based on deactivating the global constraint and posting a new global constraint with an extended set of variables. This is a general method but it requires the extended constraint to build up own data structures from scratch. Therefore, we propose to include the dynamic character into the filtering algorithm for the global constraint. In Section 4 we describe such a filtering algorithm for the dynamic `alldifferent` constraint. We conclude with a complexity study confirming that the dynamic filtering is more efficient in terms of time and space than re-posting the constraint.

## 2 Motivation

Scheduling is one of the most successful application areas of constraint programming and many global constraints were proposed for scheduling problems [7]. These global constraints are static in the sense of knowing all the constrained variables in advance. This is not a problem in the conventional formulation of scheduling problems where all the scheduled activities are known in advance and, thus, it is possible to generate all the variables and to post all the constraints before we start search for the solution. However, the static problem formulation is less appropriate for planning under resource constraints [2,4], where the activities are introduced during the course of solving. The static approach with dummy activities cannot be used there because the number of alternatives, i.e., the number of dummy activities, is very large which decreases performance (and sometimes it is even impossible to generate all the alternatives due to memory restrictions). We have observed similar difficulty when doing scheduling in complex process environments. The following section shows some examples when dynamic introduction of activities is necessary, for details and more examples see [1,2].

### 2.1 Process-dependent activities

In process production, the relations between activities are very complex and, in many cases, it is hard to express them in a mathematical way. So-called process-dependent activities are perhaps the main difficulty here, because the existence of such activity depends on allocation of other activities. Consequently, these activities cannot be introduced in advance and the global constraints that involve them cannot be posted before we know all the involved activities. Again, the static model with dummy activities cannot be used if the number of alternatives is large.

Typical examples of process-dependent activities can be identified in chemical, food, or steelmaking industries where it is necessary to insert *special set-up, cleaning, or re-heat activities* between the production activities [1,2]. In many scheduling systems, set-ups are modelled by transition times between two consecutive activities. Unfortunately, this model cannot be applied when set-up activities produce some low-quality products that must be stored or consumed by other resources[1]. In [5] another example of process-dependent activity is studied, in particular re-heat activity. Again, we need a special activity to be introduced because the re-heat activity consumes resource capacity so it must be scheduled (the re-heat activity is constrained by other activities scheduled to the same resource.

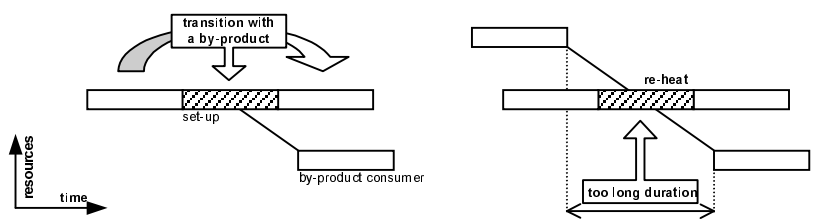

**Fig. 1.** Set-up (left) and re-heating (right). In both cases a special activity (striped rectangle) is necessary because this activity consumes resource capacity and it is connected to activities in other resources.

---

[1] Typically, these so-called by-products are not assumed in scheduling systems, which may cause problems with storing if a big quantity of by-products appear. Moreover, the by-products can be used in further production as raw material that may decrease the production cost.

## 3  Making the constraints dynamic

When constraint programming is applied to solving problems, a traditional static formulation of the problem is used: first, introduce all the variables, then post all the constraints, and finally label the variables. This schema works if we known all the objects in advance but the static schema is less appropriate when new objects (variables and constraints) appear during the process of variable labelling. This is a natural way of solving problems in Constraint Logic Programming (CLP) systems where labelling can be interleaved with introduction of new variables and constraints [3]. To implement this we need an incremental constraint solver that can accept new variables and constraints and that can remove them upon backtracking. Last In First Out (LIFO) mechanism for variable and constraint addition/retraction is used which is not complicated to implement using a stack and thanks to monotonicity of constraint systems[2]. In the rest of the paper we assume that constraints and variables are added to/removed from the constraint store in LIFO manner.

The *LIFO mechanism* can be applied to adding/removing variables to/from the dynamic global constraints as well. Note that at this stage we are not speaking about fully dynamic global constraints where the set of variables and their domains change arbitrary or this change is forced from external environment (on-line algorithms). We describe global constraints where it is possible to extend the set of variables incrementally during the course of solving (this extension is caused by the current partial solution) and to remove the variables in LIFO manner upon backtracking.

Moreover, we require the global constraint to behave *monotonically*, i.e., if we add a new variable to the constraint then the solution set does not enlarge. This is a feature of global constraints that encapsulate a set of binary constraints like the `alldifferent` constraint (such global constraints can be decomposed to the set of binary constraints over the original variables). However, there also exist global constrains which are not monotonic, like the `atleast` global constraint (at least a given number of variables take a given value). To grasp formally the monotonicity feature, we introduce monotonic global constraints.

> **Definition 1:** Let *Sol(C(X))* be a set of solutions for the global constraint *C* over the set of variables *X*. We call the **global constraint *C* monotonic** if for arbitrary variable *x* the following monotonicity property holds:
>
> $$Sol(C(X \cup \{x\})) \subseteq Sol(C(X)).$$

Note that in the above definition and in the entire paper we expect that global constraint can be defined over arbitrary (but finite) set of variables. We now show how arbitrary monotonic global constraint can be extended to a dynamic version that allows adding/removing new variables in LIFO manner.

Let $glob(\{X_1,..., X_k\})$ be a global constraint over the set of variables $\{X_1,..., X_k\}$. To capture the order of adding variables to the dynamic global constraint, we will use a list of variables $[X_1,..., X_k]$ instead of the set. When a new variable is added to the constraint then it is added to the end of this list; a variable can be removed from the constraint only if it is at the end of the list (LIFO principle). To define the dynamic version of the global constraint *glob_dyn([X$_1$,..., X$_k$])* we need the algorithms for adding a new variable to the constraint and for removing the last added variable from the constraint upon backtracking. When adding/removing the first variable to/from the dynamic constraint we can skip step 1 and step 3 respectively in the following two algorithms. Note also that the algorithms work explicitly with the stack (step 2) - this is just to illustrate the track of the algorithms (or to implement them in imperative languages), in CLP systems this is done automatically by the underlying backtracking mechanism. In fact, in the CLP system removing a variable from the global constraint is done during backtracking through the procedure for adding the variable. Thus, the programmer uses ADDVARIABLE in the CLP code (it is like posting a constraint), and when backtracking through this call, the variable is removed from the constraint. We include REMOVEVARIABLE just to illustrate the reverse process for ADDVARIABLE, this is not a real procedure that is used explicitly in the code.

---

[2]  Let *Sol(C)* be a set of solutions for the set of constraints *C*. Then for arbitrary constraint *c* the following monotonicity property holds: $Sol(C \cup \{c\}) \subseteq Sol(C)$.

**Algorithm 1:** ADDVARIABLE(glob_dyn([X$_1$,…, X$_k$]), X$_{k+1}$)

1) deactivate glob({X$_1$,…, X$_k$}), i.e., push its internal data structures to the stack and stop propagation through this constraint (if k≥1, otherwise do nothing)
2) push domains of X$_1$,…, X$_k$, X$_{k+1}$ to the stack
3) post glob({X$_1$,…, X$_k$, X$_{k+1}$}) to the constraint store

**Algorithm 2:** REMOVEVARIABLE(glob_dyn([X$_1$,…, X$_k$]))

1) remove glob({X$_1$,…, X$_k$}) from the constraint store
2) restore domains of X$_1$,…, X$_k$ from the stack
3) re-activate glob({X$_1$,…, X$_{k-1}$}), i.e. restore its internal data structures from the stack and activate propagation

The above algorithms exploit the monotonicity feature of the global constraint, i.e., if we add a new variable then the domains of already involved variables remain the same or the domains are reduced. We also expect that the variables and constraints are added/retracted in LIFO order, which is not a problem in CLP systems. Note also, that to improve efficiency we can add a set of variables together to the constraint but, then, this set is removed only from the constraint upon backtracking (it is not possible to remove a single variable from this set etc.). Still, there is hidden inefficiency:
1) we store internal representation of the global constraints after each variable addition (so we duplicate some data structures, which may increases memory consumption),
2) the dynamic constraint after adding a new variable must build its internal representation from scratch which decreases time efficiency.

The above two difficulties can be eliminated if the global constraint is implemented dynamically from beginning, i.e., if it allows adding new variables (and removing upon backtracking). We believe that at least some global constraints can accommodate such dynamic behaviour as we show in the next section for the `alldifferent` constraint.

## 4 The dynamic `alldifferent` constraint

The `alldifferent` constraint is a typical example of monotonic global constraint that exploits semantic information to achieve better domain filtering. The `alldifferent` constraint can be defined over arbitrary finite set of variables and it ensures that the variables have different values.

**Definition 2:** The `alldifferent` constraint for the set of variables $X_1,..., X_k$ is defined by the set of tuples $(d_1,..., d_k)$ such that $\forall i\ d_i \in D_i$, where $D_i$ is the domain of the variable $X_i$, and $\forall i \neq j\ d_i \neq d_j$.

$$\text{alldifferent}(X_1,..., X_k) = \{(d_1,..., d_k)\ |\ \forall i\ d_i \in D_i\ \&\ \forall i \neq j\ d_i \neq d_j\}$$

Naturally, instead of using the `alldifferent` constraint, one may post the set of binary difference constraints - for each pair of variables $X_i, X_j$ the constraint $X_i \neq X_j$ is defined. Then there is no problem when a new variable is added, we simply post the binary difference constraints between the new variable $X_{k+1}$ and all the previous variables $X_1,..., X_k$ (thus the `alldifferent` constraint is monotonic). On the other side, the `alldifferent` constraint can remove more incompatible values than the set of binary constraints as Figure 2 shows.

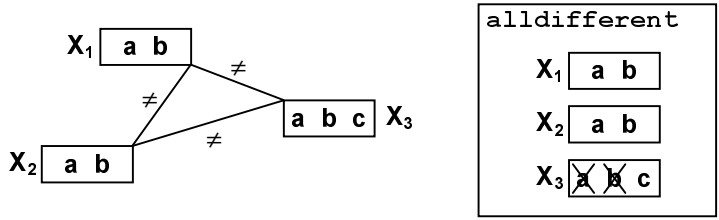

**Fig. 2.** The left constraint graph is arc-consistent so no value is found to be inconsistent. Clearly, the values *a* and *b* cannot be assigned to the variable $X_3$ so they can be removed. The `alldifferent` constraint (right) removes these values.

### 4.1 Static implementation

The efficient implementation of the `alldifferent` constraint was proposed in [6]. This implementation is based on matching theory, in particular on matching over bipartite graphs. The bipartite graph for the `alldifferent` constraint is called a value graph and it is defined in the following way:

**Definition 3:** Given an `alldifferent` constraint C, the bipartite graph $GV(C) = \langle X_C, D(X_C), E\rangle$, where $X_C$ is a set of variables in C, $D(X_C)$ is a union of domains $D_i$ for all variables in $X_C$ and $(X_i,a) \in E$ iff $a \in D_i$, is called a **value graph** of C.

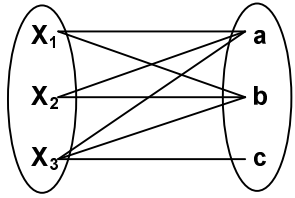

**Fig. 3.** A value graph for the `alldifferent` constraint from Figure 2.

The filtering algorithm for the `alldifferent` constraint computes the maximal matching and removes edges that are not part of any maximal matching. Note that removing an edge from the value graph is equivalent to removing a value from the variable domain. Moreover, if the maximal matching does not cover all the variables from the `alldifferent` constraint then the constraint fails (it is not possible to find a consistent labelling of variables).

**Algorithm 3:** ALLDIFF-INITIALISATION(C)

```
1) build G = ⟨X_C, D(X_C), E⟩,
2) M(G) ← ComputeMaximumMatching(G)
3) if |M(G)|<| X_C | then return false
4) RemoveEdgesFromG(G,M(G))
5) return true
```

The procedure REMOVEEDGESFROMG deletes every edge that does not belong to any matching which covers $X_C$. Such edges are found by exploring even alternating paths and cycles. The description of linear algorithm can be found in [6].

Naturally, the constraint systems consist of more than single `alldifferent` constraint and the other constraints may reduce the domains of variables from $X_C$ as well. We can repeat the ALLDIFF-INITIALISATION procedure each time a domain of a variable from $X_C$ is changed (a value is deleted) but we can do better by using the fact that before the deletion, a matching which covers $X_C$ is known. The algorithm proposed in [6] uses a function MATCHINGCOVERINGX(G, $M_1$, $M_2$), which computes a matching $M_2$, which covers $X_C$, from a matching $M_1$, which is not maximal. If no such matching exists then the procedure returns *false*. We present here a slightly simplified version of the algorithm from [6] (we expect that when an edge (*X*,*a*) is removed from the graph then the value *a* is not the only value for the variable *X*, otherwise the constraint that causes such reduction has already failed). The algorithm gets the value graph G, the maximum matching M(G), and the list of edges ER to delete as input. It returns *false* if there is no maximum matching which covers $X_C$, and it returns *true* otherwise and deletes every edge that does not belong to any matching which covers $X_C$.

**Algorithm 4:** ALLDIFF-PROPAGATION(G,M(G),ER)

```
1)   computeMatching ← false
2)   for each e∈ER do
3)       if e∈M(G) then
4)           M(G) ← M(G) - {e}
5)           computeMatching ← true
6)       remove e from G
7)   if computeMatching then
8)       if ¬MATCHINGCOVERINGX(G,M(G),M') then
9)           return false
10)      else
11)          M(G) ← M'
12)  REMOVEEDGESFROMG(G,M(G))
13)  return true
```

### 4.2 Dynamic extension

As we mentioned in Section 3, by a dynamic global constraint we mean a constraint whose set of variables can be extended and shrunk in LIFO manner. Thus, in the following we concentrate on the algorithm for adding a new variable to the constraint. We expect that the variable can be removed upon backtracking only (and the underlying system restores the data structures to the state just before the variable was added).

Remind that the `alldifferent` constraint is monotonic so we can add a new variable (or a set of variables) to the constraint without extending the solution set. Moreover, we can incrementally extend the value graph to get a value graph for the constraint with more variables. In fact we are extending the set $X_C$ of variables, the set $D(X_C)$ of values (perhaps), and the set E of edges. The new edges connect only the new variables with values (old or new); there is no new edge connecting any old variable with any value (we are not extending the domains of the old variables so we keep monotonicity).

**Definition 4:** Let $\langle X, D(X), E \rangle$ be a value graph of the `alldifferent` constraint C(X) for the set X of variables. Then the bipartite graph $\langle X \cup Y, D(X \cup Y), E' \rangle$ where Y is set of added variables and $E' = E \cup \{(X_i,a) | X_i \in Y \text{ \& } a \in D_i, \}$ is called an **extended value graph** of $C(X \cup Y)$ for the extended set $X \cup Y$ of variables.

The important thing is that the extension of the variable set does not influence the decisions taken before. In particular, if any edge is deleted from the original value graph (because it does not belong to any matching which covers X) then this edge does not belong to any matching which covers $X \cup Y$ (the constraint is monotonic).

This observation is used in the algorithm that updates the value graph after adding new variables. It means that we can incrementally update the value graph after adding new variables instead of computing a new value graph from scratch.

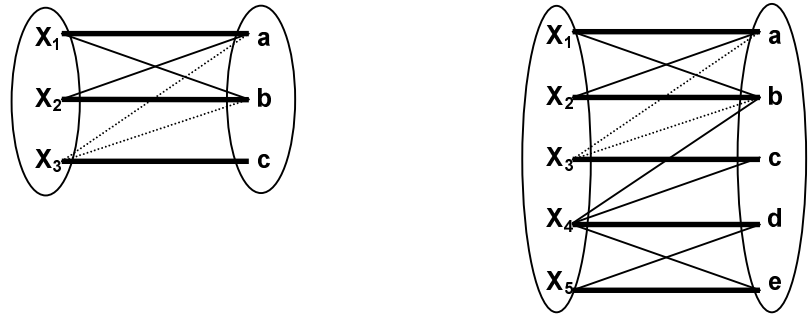

**Fig. 4.** Edges deleted (dotted edges) from the original value graph (left) do not belong to the extended value graph (right). Bold edges are the maximum matching edges.

After extending the original value graph by adding new edges, the former maximum matching does not cover the extended set X∪Y of variables. We can use the function MATCHINGCOVERINGX to extend the matching which covers X to a matching which covers X∪Y. If there exists a maximum matching which covers X∪Y then the edges that do not belong to any maximum matching are removed using REMOVEEDGESFROMG and the algorithm returns *true*, otherwise it returns *false*.

**Algorithm 5:** ALLDIFF-UPDATE(G,M(G),EA)

```
1) for each e∈ EA do
2)     add e to G
3) if ¬MATCHINGCOVERINGX(G,M(G),M') then
4)     return false
5) else
6)     M(G) ← M'
7) REMOVEEDGESFROMG(G,M(G))
8) return true
```

Removing a variable from the alldifferent constraint is done upon backtracking only so we do not present a special algorithm for this. The standard mechanism with a stack to recover data structures can be used there. Initialisation and propagation algorithms for the dynamic alldifferent constraint are the same as for the static version, i.e., Algorithm 3 and Algorithm 4.

# 5   Complexity Analysis

Besides the dynamic filtering for the `alldifferent` constraint (presented in Section 4.2) we can use a generic dynamisation technique proposed in Section 3. Let us now compare the complexity of these two methods for making the `alldifferent` constraint dynamic. We will use the basic complexity results from [6].

Assume that $p$ is a number of variables in the constraint and $d$ is a number of all different values in the domains of these variables (a size of the union of the domains). Thus $p+d$ is a number of vertices in the value graph. Let $m$ be a number of edges in the value graph, clearly $m \leq pd$ (the value graph is a bipartite graph).

**Space complexity.** To represent the `alldifferent` constraint we need keeping its value graph. The graph can be represented by the set of its edges so the space complexity is $O(m)$ i.e., $O(pd)$. If a generic dynamisation technique is used then a new value graph is introduced (and the previous value graph is kept in memory) after adding a new variable. Thus, space complexity of adding a new variable is $O(pd)$.

In case of dynamic filtering algorithm, only new edges are added to the value graph when new variable arrives. These edges can only connect the new variable with the values, so maximal number of new edges is $d$. Thus, space complexity is $O(d)$, $p$-times better than generic dynamisation.

**Time complexity.** According to [6], time complexity of AllDiff-Initialisation is $O(dp\sqrt{p})$. Thus, time complexity of adding a new variable using the generic dynamisation technique is $O(dp\sqrt{p})$ as well (in fact, we are adding a new constraint).

Time complexity of RemoveEdgesFromG is $O(m+p+d)$ [6] so we can use $O(pd)$ instead. Time complexity of MatchingCoveringX is $O(m\sqrt{k})$, where $k$ is a number of edges missing in the maximal covering (these edges must be added to find a covering which covers all the variables) [6]. Note that if we are adding a new variable then exactly one edge is missing in the maximal covering - the edge connecting the new variable with some value. Consequently, time complexity of MatchingCoveringX called in AllDiff-Update is $O(m)$, that is $O(pd)$. Together, time complexity of adding a new variable using AllDiff-Update is $O(pd)$, $\sqrt{p}$-times better than generic dynamisation.

However, note that if we know the variables in advance, it is more efficient to add them all together, time complexity is $O(dp\sqrt{p})$, than adding them incrementally (one by one), time complexity is $O(dp^2)$. Our algorithm supports adding more variables together in an efficient way with time complexity $O(dp\sqrt{k})$, where $k$ is a number of added variables.

Time complexity of propagation of deletions (a value is removed from the domain of a variable by another constraint) through the constraint is $O(pd)$ [6], same for both methods of dynamisation. Nevertheless, note that some systems do not support deactivating the constraint from the constraint store - the constraint is removed upon backtracking only. Consequently, when generic dynamisation is used then all the constraints posted so far are active and propagation is duplicated, which decreases overall efficiency.

|                                              | SPACE   | TIME           |
|----------------------------------------------|---------|----------------|
| ADDVARIABLE(ALLDIFF) *generic dynamisation*  | $O(pd)$ | $O(dp\sqrt{p})$ |
| ALLDIFF-UPDATE *dynamic all-different*       | $O(d)$  | $O(dp)$        |

**Table 1.** Comparison of time and space complexity of adding a variable for the generic dynamisation and the dynamic implementation of the `alldifferent` constraint.

## 6  Conclusions

Wider and wider application area of constraint programming brings new challenges to constraint satisfaction algorithms. Dynamisation of global constraints, i.e., extending a global constraint to work with the changing set of variables, is one of the desirable features. In the paper we give some examples when a dynamic approach to constraint satisfaction problem is necessary. We argue here for using dynamic global constraints which provide good domain filtering and which can adopt new variables if necessary. We showed a general mechanism for dynamisation of global constraints and we presented a dynamic version of the `alldifferent` constraint. We also proved that the dynamic version of the `alldifferent` constraint is more efficient in terms of time and space than re-posting a constraint each time a new variable arrives (generic dynamisation).